\documentstyle[aps,epsfig]{revtex}
\def \blankline{\vspace{0.4 cm}}

\begin {document}
\draft

\title {\bf Upper limit on the 
prompt muon flux derived from the LVD underground experiment}
\blankline
\author{
{\bf The LVD Collaboration}\\
\vspace{0.3cm}
M.Aglietta$^{16}$, B.Alpat$^{13}$, E.D.Alyea$^{7}$, P.Antonioli$^{1}$,
G.Badino$^{16}$, G.Bari$^{1}$, M.Basile$^{1}$,
V.S.Berezinsky$^{10}$, F.Bersani$^{1}$, M.Bertaina$^{16}$, 
R.Bertoni$^{16}$, G.Bonoli$^{1}$, A.Bosco$^{2}$,
G.Bruni$^{1}$, G.Cara Romeo$^{1}$,
C.Castagnoli$^{16}$, A.Castellina$^{16}$, A.Chiavassa$^{16}$,
J.A.Chinellato$^{3}$, 
L.Cifarelli$^{1,\dagger}$, F.Cindolo$^{1}$,
G.Conforto$^{17}$,
A.Contin$^{1}$, V.L.Dadykin$^{10}$,
A.De Silva$^{2}$, M.Deutsch$^{8}$, P.Dominici$^{17}$,
L.G.Dos Santos$^{3}$, L.Emaldi$^{1}$,
R.I.Enikeev$^{10}$, F.L.Fabbri$^{4}$, W.Fulgione$^{16}$,
P.Galeotti$^{16}$, C.Ghetti$^{1}$,
P.Ghia$^{16}$, P.Giusti$^{1}$, R.Granella$^{16}$, F.Grianti$^{1}$,
G.Guidi$^{17}$,
E.S.Hafen$^{8}$, P.Haridas$^{8}$, G.Iacobucci$^{1}$, N.Inoue$^{14}$,
E.Kemp$^{3}$, F.F.Khalchukov$^{10}$, E.V.Korolkova$^{10}$,
P.V.Korchaguin$^{10}$, V.B.Korchaguin$^{10}$, 
V.A.Kudryavtsev$^{10,\ddagger}$,
K.Lau$^{6}$, M.Luvisetto$^{1}$, G.Maccarone$^{4}$,
A.S.Malguin$^{10}$, R.Mantovani$^{17}$,
T.Massam$^{1}$,
B.Mayes$^{6}$, A.Megna$^{17}$, C.Melagrana$^{16}$,
N.Mengotti Silva$^{3}$,
C.Morello$^{16}$, J.Moromisato$^{9}$,
R.Nania$^{1}$,
G.Navarra$^{16}$, L.Panaro$^{16}$,
L.Periale$^{16}$, A.Pesci$^{1}$, P.Picchi$^{16}$, L.Pinsky$^{6}$,
I.A.Pless$^{8}$, J.Pyrlik$^{6}$, V.G.Ryasny$^{10}$,
O.G.Ryazhskaya$^{10}$, O.Saavedra$^{16}$, K.Saitoh$^{15}$,
S.Santini$^{17}$, G.Sartorelli$^{1}$, 
M.Selvi$^{1}$,
N.Taborgna$^{5}$,
V.P.Talochkin$^{10}$,
J.Tang$^{8}$, G.C.Trinchero$^{16}$, S.Tsuji$^{11}$, A.Turtelli$^{3}$,
I.Uman$^{13}$, P.Vallania$^{16}$, G. Van Buren $^{8}$, 
S.Vernetto$^{16}$,
F.Vetrano$^{17}$, C.Vigorito$^{16}$, E. von Goeler$^{9}$,
L.Votano$^{4}$, T.Wada$^{11}$,
R.Weinstein$^{6}$, M.Widgoff$^{2}$,
V.F.Yakushev$^{10}$, I.Yamamoto$^{12}$,
G.T.Zatsepin$^{10}$, A.Zichichi$^{1}$
}
\vspace{0.3cm}
\address{
\noindent $^{1}${\it University of Bologna and INFN-Bologna, Italy}\\
$\,^{2}${\it Brown University, Providence, USA}\\
$\,^{3}${\it University of Campinas, Campinas, Brazil}\\
$\,^{4}${\it INFN-LNF, Frascati, Italy}\\
$\,^{5}${\it INFN-LNGS, Assergi, Italy}\\
$\,^{6}${\it University of Houston, Houston, USA}\\
$\,^{7}${\it Indiana University, Bloomington, USA}\\
$\,^{8}${\it Massachusetts Institute of Technology, Cambridge, USA}\\
$\,^{9}${\it Northeastern University, Boston, USA}\\
$^{10}${\it Institute for Nuclear Research, Russian Academy of
Sciences, Moscow, Russia}\\
$^{11}${\it Okayama University, Okayama, Japan}\\
$^{12}${\it Okayama University of Science, Okayama, Japan}\\
$^{13}${\it University of Perugia and INFN-Perugia, Italy}\\
$^{14}${\it Saitama University, Saitama, Japan}\\
$^{15}${\it Ashikaga Institute of Technology, Ashikaga, Japan}\\
$^{16}${\it Institute of Cosmo-Geophysics, CNR, Torino, University
of Torino and} \\
\indent {\it INFN-Torino, Italy}\\
\noindent $^{17}${\it University of Urbino and INFN-Firenze, Italy}\\
\vspace{0.3cm}
$^{\dagger}${\it now at University of Salerno and INFN-Salerno, Italy}\\
$^{\ddagger}${\it now at University of Sheffield, United Kingdom}
}
\maketitle

\begin{abstract}
We present the analysis of the muon events with all muon multiplicities
collected during 21804
hours of operation of the first LVD tower. The measured depth -- angular
distribution of muon intensities has been used to obtain the 
normalization 
factor, $A$, the power index, $\gamma$, of the primary all-nucleon 
spectrum and the ratio, $R_c$, of prompt muon flux to that of 
$\pi-$mesons -- the main parameters which determine the spectrum of cosmic ray
muons at the sea level. The value of $\gamma=2.77 \pm 0.05$ 
(68$\%$ C.L.) and
$R_c<2.0 \cdot 10^{-3}$ (95$\%$ C.L.) have been obtained.
The upper limit to the prompt muon flux favours the
models of charm production based on QGSM and the dual parton model.
\end{abstract}
\pacs{PACS numbers: 13.85.Tp, 13.85.-t, 96.40.Tv, 96.40.-z}

\section { Introduction }

The depth -- angular distribution of muon intensity measured in an
underground experiment is closely related to the muon energy spectrum
at surface. Assuming the muon survival probabilities are well known for
every depth and every muon energy at surface, the analysis of the
measured depth -- zenith angle distribution of intensity allows us to
evaluate the parameters of the muon spectrum at the sea level, i.e.
the normalization constant, the power index of the primary all-nucleon
spectrum, $\gamma$, 
and the prompt muon flux from the decay of charmed particles
produced together with pions and kaons in the high-energy hadron-nucleus
interactions.

Among these characteristics the value of prompt muon flux attracts
a particular interest. It can be evaluated from the zenith-angle
distributions of muon intensities, measured at various muon energies or
various depths. The fraction of prompt muons cannot be estimated from 
the muon
energy spectrum or depth-intensity curve measured at one zenith angle
because the same effect can be produced either by the prompt muons, or
the decrease of $\gamma$, or both.

The charmed particles are produced together with pions and kaons
in the collisions of primary cosmic rays with air nuclei. They have
such short
live times that they decay immediately (if their energy is less than
1000 TeV) into muons and other particles. Thus, for them there is no
competition between interaction and decay, and the prompt muon energy
spectrum has almost the same slope as the primary spectrum. Due to
the rise of the charm production cross-section in the energy range
100--1000 TeV, the power index of the prompt muon spectrum, $\gamma_c$, 
can be little lower than $\gamma$.
However, possible sca\-ling violation in the fragmentation region 
can increase the value of $\gamma_c$.
Due to the absence of the competition between
interaction and decay of charmed particles the zenith-angle
distribution
of prompt muons is almost flat, comparing with the
$\sec \theta$ - distribution of the conventional muons (from the
decay of
pions and kaons). This allows to estimate the fraction of prompt muons
by analysing the zenith-angle distribution of muon intensities.

\indent Numerous calculations of the prompt muon flux were done
(see, for example,
\cite{Elbert,Cast,Inazawa,Volkovac,Zas,Bugaev,Forti,Thunman}).
Different models give the prompt muon fluxes which vary by 2
orders of magnitude. This is due to the uncertainties in the charm
production
cross-section, $\sigma_c$, $x$-distribution of charmed particles
($x=E_c/E_0$),
produced in pA-collisions, and the branching ratio of charmed particle 
decay into mu\-ons. The most uncertain parameter, that results in 
the large dispersion of the predicted prompt muon flux, 
is the $x$-distribution of produced charmed particles
in the fragmentation region, important for the charm-produced cosmic-ray
muons. This distribution at high energies cannot be measured precisely
at accelerators which give the information only about small $x$. Thus,
to check the models of the charm production, the experiments with
cosmic-ray muons at high energies are useful.

The search for the prompt muon flux was done with several detectors
located at the surface and underground (see, for example,
\cite{KGF,NUSEXc,Baksanc,MSU}). In practice,
it is convenient to express the prompt muon flux in terms of the
ratio, $R_c$, of prompt muon flux to that of pions at vertical.
Since the slope of the prompt muon spectrum is close to that of pion
spectrum, the ratio $R_c$ is almost constant for all muon energies
available in the existing experiments.
The experimental data, collected up to now, show a large variation
of $R_c$ (from $0$ to $4 \cdot 10^{-3}$).

In a previous paper \cite{LVD} we have presented our
measurement of the single muon `depth -- vertical intensity' curve 
and the evaluation of the
power index of the meson spectrum in the atmosphere using the
`depth -- vertical intensity' relation for single muons. 
Here we present the analysis of all muon sample which include the muon
events with all multiplicities.
The muon survival probabilities, used to
obtain the value of $\gamma$ in \cite{LVD}, have been presented
in \cite{LVDR}. They have been calculated using the muon interaction
cross-sections from \cite{BBb,BBn,KP}. After the publication of these
results, new calculation of the cross-section of muon bremsstrahlung
and of the corrections to the knock-on electron production cross-section
have been done \cite{KKP}. In the present analysis we have taken into
account the corrections proposed in \cite{KKP} and we have estimated
the uncertainties of $\gamma$ due to the uncertainties of the
cross-sections used to simulate the muon transport through the rock.
In this paper we present a more detailed evaluation of the
characteristics of the muon spectrum at the sea level, including the
ratio of the prompt muon flux to that of pions, using the
depth -- zenith angle distributions of muon intensities
($I_{\mu}(x,\theta)$) measured with LVD in the underground 
Gran Sasso Laboratory.
The analysis is based on an increased statistics comparing
with the previous publications.
The `depth -- vertical intensity' relation for all muon sample and
its analysis are presented in a separate paper~\cite{lvdlvd}.

In Section 2 the detector and the procedure of data processing
are briefly described. In Section 3
the results of the analysis of the muon intensity
distribution ($I_{\mu}(x,\theta)$) are
presented. In Section 4 we discuss our
results in comparison with the data of other experiments and
theoretical expectations. Section 5 contains the conclusions.

\section { LVD and data processing }

The LVD (Large Volume Detector) experiment is located in the
underground Gran Sasso Laboratory at a minimal depth of about
3000 hg/cm$^2$.
The LVD will consist of 5 towers. The 1st tower is running since
June, 1992,
and the 2nd one - since June, 1994. The data presented here were
collected with the 1st LVD tower during 21804 hours of live time.

The 1st LVD tower contains 38 identical modules \cite{LVDa}.
Each module con\-sists of 8 scintillation counters and 4 layers of 
limited
streamer tubes (tracking detector) attached to the bottom and to one
vertical side of the supporting structure. A detailed description of the
detector was given in \cite{LVDa}. One LVD tower has the dimensions of
$13 \times 6.3 \times 12$ m$^3$.

The LVD measures the atmospheric muon intensities from
3000 hg/cm$^2$ to more than 12000 hg/cm$^2$ (which
correspond to the median muon energies at the sea level from 1.5 TeV to
40 TeV) at the zenith angles from $0^o$ to $90^o$
(on the average, the larger depths correspond to higher zenith angles).

We have used in the analysis the muon events with all multiplicities,
as well as the sample of single muons. Our basic results have been
obtained with all muon sample. This sample contains about 2 millions of
reconstructed muon tracks.

The acceptances for each angular bin have been calculated using
the simulation of muons passing through LVD taking into account
muon interactions with the detector materials and the detector
response. The acceptances for both single and multiple muons were
assumed to be the same.

As a result of the data processing the angular distribution
of the number of detected muons
$N_{\mu}(\phi, \cos \theta)$ has been obtained. The angular bin
width $1^o \times 0.01$ has been used. The analysis refers to 
the angular bins for which
the efficiency of the muon detection and track reconstruction is greater
than 0.03. We have excluded from the analysis the angular bins with
a large variation of depth.

The measured $N_{\mu}(\phi, \cos \theta)$-distribution has been
converted to the depth -- angular distribution of muon intensities,
$I_{\mu}(x,\cos \theta)$, using the formula:

\begin{equation}
I_{\mu}(x_m,\cos \theta_i)=
{{\sum_{j} N_{\mu}(x_m(\phi_j),\cos \theta_i)} \over
{\sum_{j} (A(x_m(\phi_j),\cos \theta_i) 
\epsilon(x_m(\phi_j),\cos \theta_i) \cdot \Omega_{ij} \cdot T)}}
\label{depth-angle}
\end{equation}

where the summing up has been done over all angles $\phi_j$ 
contributing to the depth $x_m$; $A(x_m(\phi_j),\cos \theta_i)$ is 
the cross-section of the
detector in the plane perpendicular to the muon track at the angles
($\phi_j,\cos \theta_i$); $\epsilon(x_m(\phi_j),\cos \theta_i)$ is the
efficiency of muon detection and reconstruction; $\Omega_{ij}$ is the solid
angle for the angular bin, and $T$ is the live time.
We have chosen the depth bin width increasing with the
depth to have comparable statistics at all depth bins from 3 to 10
km w.e.. Thus, the depth bin width increases from about 100 m w.e.
at 3000 m w.e. to more than 500 m w.e. at about 10000 m w.e.
The muon intensities have been converted to the middle points of the
depth bins taking into account the predicted depth -- intensity relations
for different zenith angles (we have used the parameters of the muon
spectrum at sea level which fit well the `depth -- vertical muon
intensity' relation measured by LVD \cite{LVD}). The angular bin width
has been taken equal to $\Delta(\cos \theta)=0.025$. The conversion to the
middle points of the angular bins has been done according to the
predicted angular dependence for muons from pion and kaon decay.
However, due to the small angular bins this conversion does not
change angular distributions.

\section { Analysis of the
depth -- zenith angle distribution of muon intensity measured by LVD }

\begin{figure}[htb]
\epsfig{figure=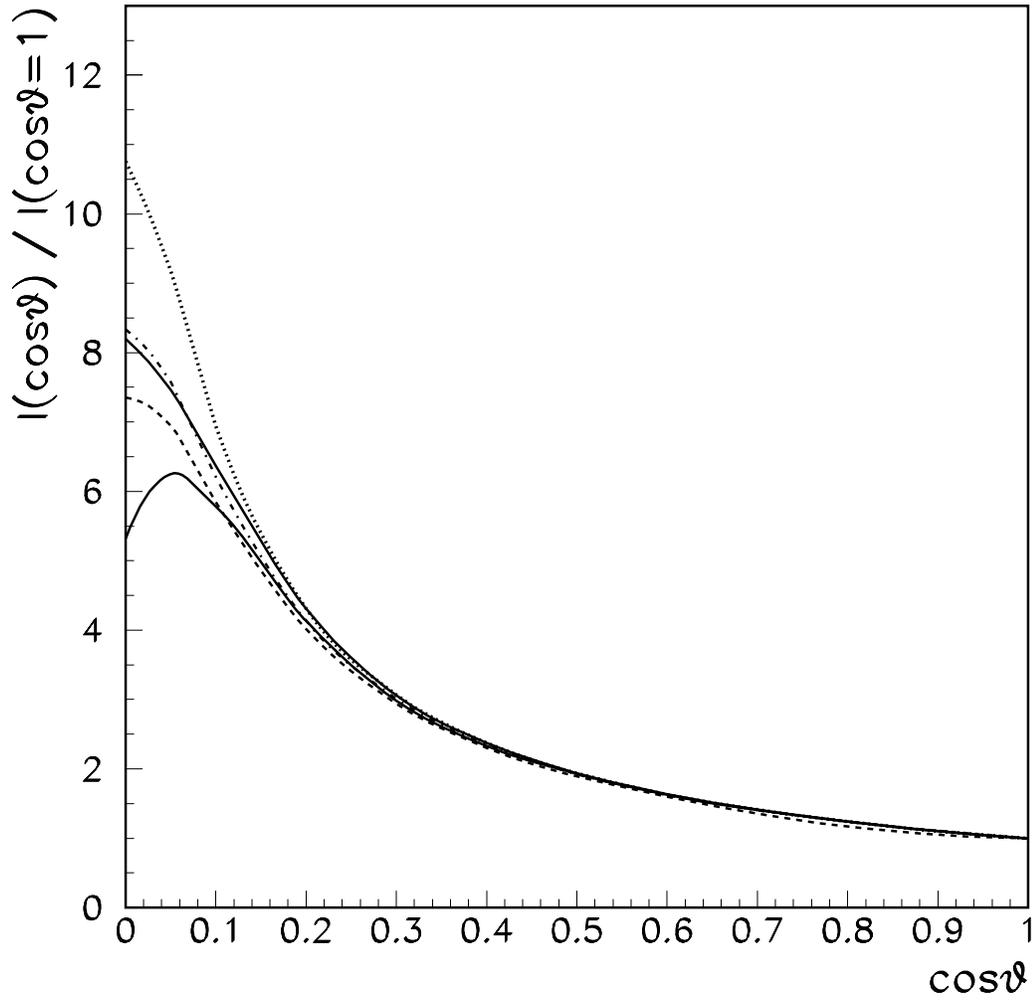,height=15cm}
\caption{ Ratio of muon intensity at $\cos\theta$ to that at 
$\cos\theta=1$ for 10 TeV muons at sea level versus cosine of
zenith angle $\cos\theta$ calculated using different formulae:
dotted curve -- eq. (3) 
with $\cos\theta^{\star}$ from
Earth curvature with scale height of 17 km;
dash-dotted curve -- eq. (3) with 
$\cos\theta^{\star}$ from
Earth curvature with scale height of 32 km;
upper solid curve -- eq. (3) with 
$\cos\theta^{\star}$ from [21];
dashed curve -- calculations of [22];
lower solid curve -- original formula from [21,23].}
{\label{angle}}
\end{figure}

The data analysis has included the procedure of fitting of the measured
depth -- zenith angle distribution of muon intensity with the
distributions calculated using the known muon survival probabilites
(see \cite{LVD,LVDR}, and references therein)
modified for a
new muon bremsstrahlung cross-section \cite{KKP}
and muon spectrum
at sea level with three free parameters: normalization constant, $A$,
power index of primary all-nucleon spectrum, $\gamma$, and the ratio
of prompt muons to pions, $R_c$. The depth -- angular distributions of
muon intensity have been calculated using the equation:

\begin{equation}
I_{\mu}(x,cos\theta)=\int_0^{\infty} P(E_{\mu 0},x)
\cdot {{d I_{\mu 0}(E_{\mu 0},\cos\theta)}\over{d E_{\mu 0}}}
\cdot d E_{\mu 0},           \label{depth-angle calculated}
\end{equation}

where $P(E_{\mu 0},x)$ is the probability for muon with
an initial energy $E_{\mu 0}$ at sea level to survive at the
depth $x$ in Gran Sasso rock, and
${{d I_{\mu 0}(E_{\mu 0},\cos\theta)}\over{d E_{\mu 0}}}$ is the
muon spectrum at sea level which has been taken according to
\cite{Gaisser}:

\begin{eqnarray}
{{d I_{\mu 0} (E_{\mu 0}, \cos\theta)}\over{d E_{\mu 0}}}
& = &
A \cdot 0.14 \cdot E_{\mu 0}^{-\gamma} \nonumber \\
& \times &
\left({{1}\over{1+{{1.1 E_{\mu 0} \cos\theta^{\star}}\over{115 GeV}}}}+
{{0.054}\over{1+{{1.1 E_{\mu 0} \cos\theta^{\star}}\over{850 GeV}}}}
+ R_c\right)
\label{Gaisser spectrum}
\end{eqnarray}

where the values of $cos\theta$ have been substituted by
$cos\theta^{\star}$ which have been taken from either \cite{Volkovacos}
or a simple consideration of the curvature of the Earth atmosphere.
In a search for a small contribution of prompt muons it is necessary
to know precisely the angular dependence of conventional (from pion and
kaon decay) muon intensity at all energies of interest.
In \cite{Volkovacos} $cos\theta^{\star}=
E_{\pi,K}^{cr}(cos\theta=1)/E_{\pi,K}^{cr}(cos\theta)$, where
$E_{\pi,K}^{cr}$ are the critical energies of pions and kaons.
$cos\theta^{\star}$ can be understood also as
the cosine of zenith angle of muon direction at the height of
muon production. The height of muon production increases from
17 km at $cos\theta=1$ to about 32 km at $cos\theta=0$.
We have found that the
values of $cos\theta^{\star}$ depend on the model of the atmosphere in the
range of $cos\theta=0-0.3$. 
In Figure \ref{angle} we present the predicted angular dependences of 
conventional muon intensities
at the energy of 10 TeV. As can be seen, all curves almost coincide at 
$cos\theta=0.3-1$. However, there is a large spread of functions at
$cos\theta=0-0.3$. The calculations using eq. (\ref{Gaisser spectrum})
with $cos\theta^{\star}$ from \cite{Volkovacos} (upper solid curve)
or the treatment
of the Earth curvature with a muon production height of 32 km
(dash-dotted curve),
as well as the results of \cite{Lipari} (dashed curve) give
quite similar results at all $cos\theta$, while the original
calculations of \cite{Volkovacos,Volkovasp} (lower solid curve) 
and the treatment
of the Earth curvature with a muon production height of 17 km
(dotted curve)
are far below or above other curves at small $cos\theta$.
To be independent of the model we have
restricted the range of $cos\theta$ used in the analysis to 0.3 -- 1.
This increases the statistical error of the results decreasing at the same
time the systematical uncertainty related to model used.
This also reduces the sensitivity of the experiment to small
values of $R_c$.
We note that the uncertanties in the rock thickness and rock density are 
high enough at small $cos\theta$. Moreover, large derivative of the column 
density with angle together with muon scattering effect lead to the high
uncertainties of the muon flux. This also justifies our decision 
to restrict the range of zenith angles used in the analysis.

We have added to the original formula  of \cite{Gaisser} the term
$R_c$, which is the ratio of prompt muons to pions.
Here it has been assumed that
the power index of the prompt muon spectrum
is equal to that of primary spectrum.
Really, due to rapid rise of charm production cross-section and
the possible scaling violation in the fragmentation region,
the prompt muon spectrum may have the power index, $\gamma_c$, 
different from
$\gamma$. But the value of $\gamma_c$ depends on the
model of charm production. To be independent of the models
we have used at the first approximation the assumption:
$\gamma_c$=$\gamma$.
The full formula has been multiplied by the additional normalization 
constant $A$ which has
been considered as a free parameter together with $\gamma$ and $R_c$.

\begin{figure}[htb]
\epsfig{figure=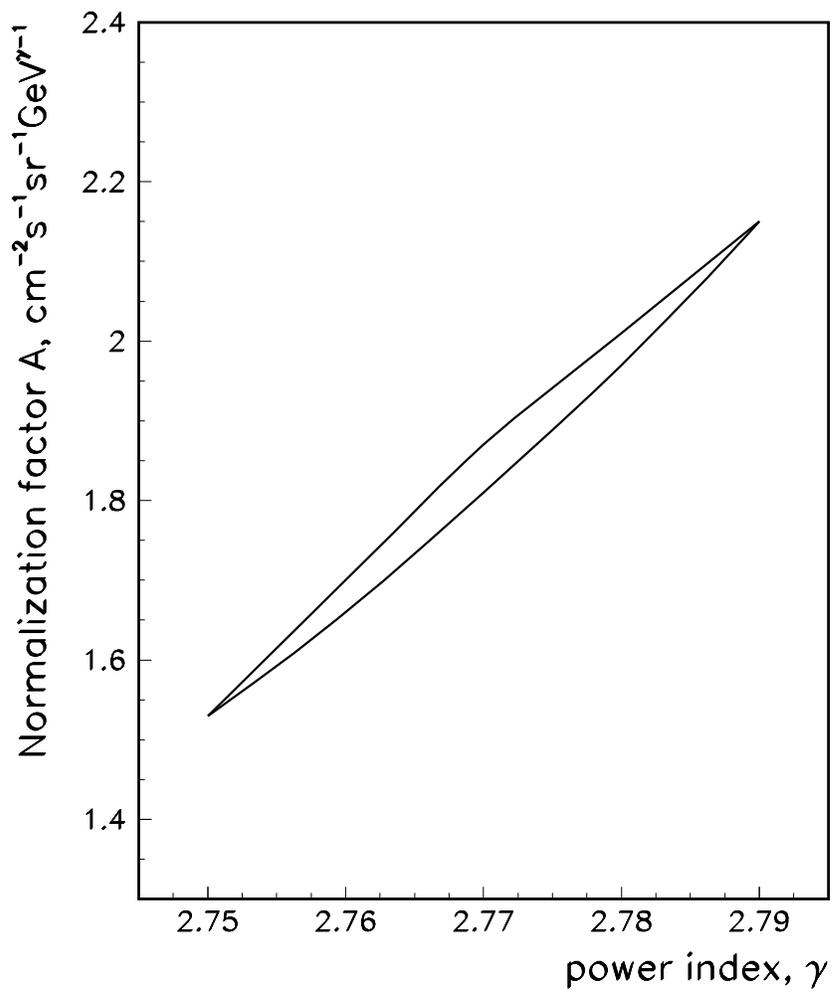,height=15cm}
\caption{ Contour plot of allowed region in $A-\gamma$ -- plane
showing strong correlation between the parameters.}
{\label{a-gamma}}
\end{figure}

\begin{figure}[htb]
\epsfig{figure=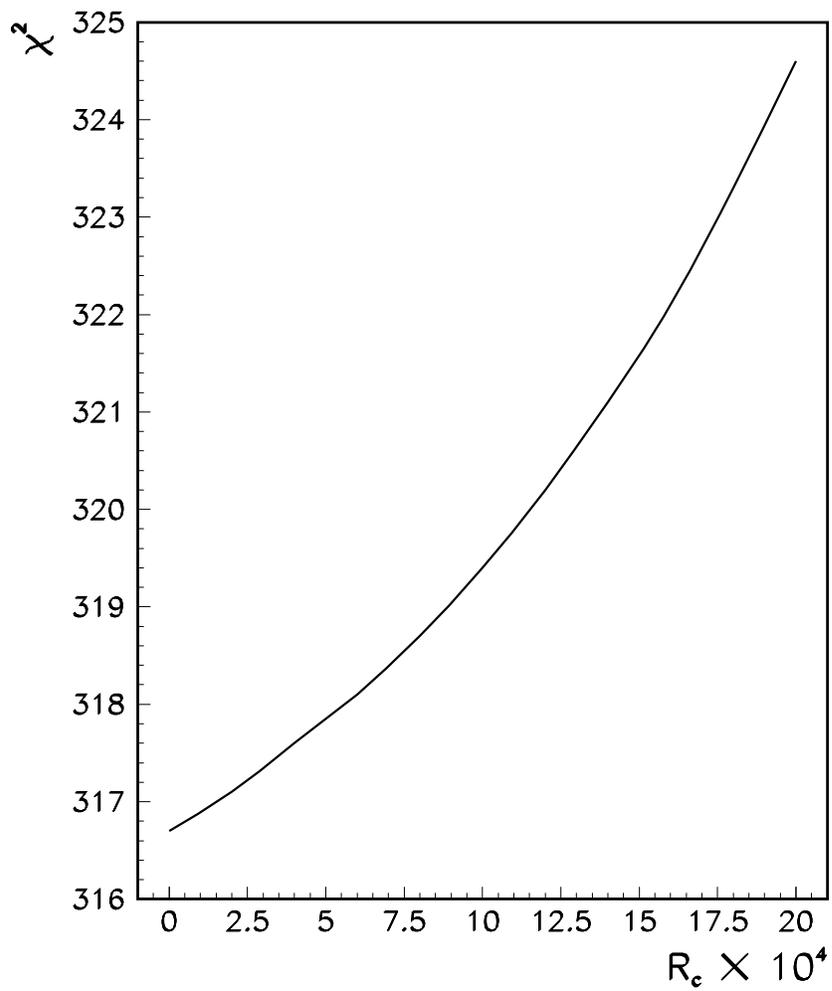,height=15cm}
\caption{ Dependence of $\chi^2$ on the ratio of prompt muon
flux to that of pions.}
{\label{chi2}}
\end{figure}

As a result of the fitting procedure we have obtained the values of
the free parameters: $A=1.84 \pm 0.31$, $\gamma=2.77 \pm 0.02$ and the
upper limit on $R_c \leq 2 \cdot 10^{-3}$. Here and hereafter we
present the errors at 68\% confidence level (C.L.) and the upper limits at
95\% C.L. The value of $\chi^2$ is equal to 316.7 for 330 degrees of freedom.
The estimates of the parameters $A$ and $\gamma$ are strongly
correlated. The larger the value of $\gamma$ is, the larger
the normalization factor $A$ should be. Figure \ref{a-gamma} shows
the contour plot of allowed region in $A-\gamma$ -- plane.
The dependence of $\chi^2$ on $R_c$ is presented in Figure \ref{chi2}
which was used to obtain an upper limit on $R_c$.
The errors of the parameters include both
statistical and systematic uncertainties. The latter one takes into account
the possible uncertainties in the depth and local density,
but does not take into account the uncertainty in the cross-sections used
to simulate the muon transport through the rock. If we add the
uncertainty in the muon interaction cross-sections, the error of $\gamma$
will increase from 0.02 to 0.05 (for the discussion about the uncertainty
due to different cross-sections see \cite{MUSIC}).
This uncertainty, however, does not influence the upper limit on $R_c$.
We note that the energy in eq. (\ref{Gaisser spectrum}) is expressed
in GeV and the intensity is expressed in cm$^{-2}$ s$^{-1}$ sr$^{-1}$.
If we restrict our analysis to the depth range 5 -- 10 km w.e., we 
obtain the following values of parameters: $A=1.6^{+0.8}_{-0.6}$,
$\gamma=2.76 \pm 0.06$ and $R_c \leq 3 \cdot 10^{-3}$.

\begin{figure}[htb]
\epsfig{figure=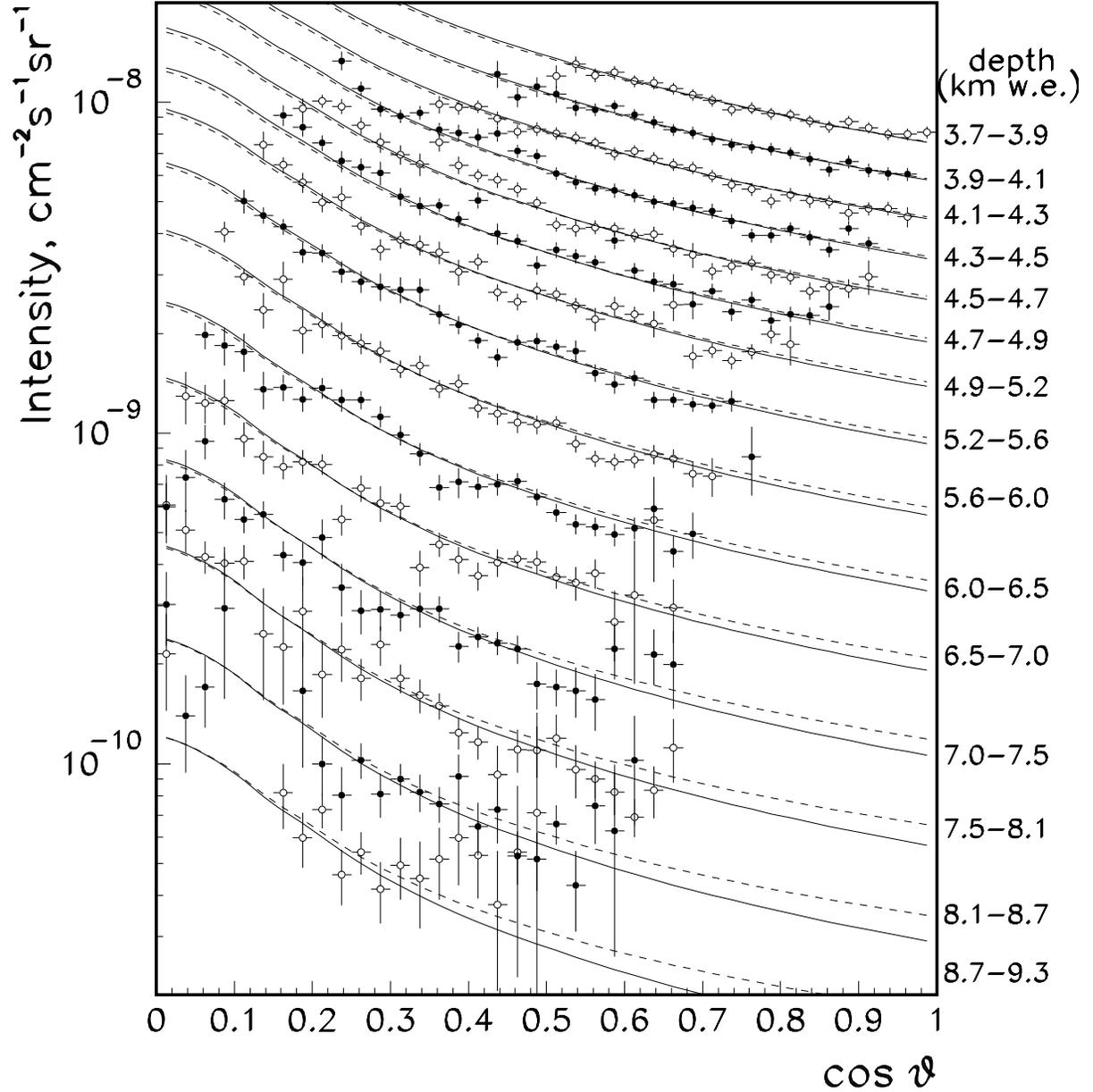,height=19cm}
\caption{ The dependence of the muon intensity
on the zenith angle for the depth bins of most interest
in the analysis and for all zenith angles.
The data have been converted to the middle points
of depth and angular bins.
Solid curve - calculation with $\gamma = 2.77 $ and $R_c = 0$
(best fit to the LVD data in the whole depth range, see eq. (3));
dashed curve - calculation with $\gamma = 2.77 $ and
$R_c = 2 \cdot {10^{-3}}$ (LVD upper limit).
The absolute normalization of both sets of calculations
has been done independently using the fitting procedure.
The error bars
include both statistical and systematic uncertainties.}
{\label{depth_angle}}
\end{figure}

The angular distributions of muon intensities for 
depth ranges of interest
are presented in Figure \ref{depth_angle} together with calculations with
$R_c=0$ (best fit -- solid curve) and $R_c=2 \cdot 10^{-3}$
(upper limit -- dashed curve). 
The normalizations of both cal\-cu\-la\-tions have
been done independently using the fitting procedure. 
The data at all zenith angles are shown but the analysis was restricted 
to the range $0.3<cos\theta<1$.
The error bars
show both statistical and systematic uncertainties.
The calculated distributions
have been obtained using the eq. (\ref{Gaisser spectrum}) and the
values of $cos\theta^{\star}$ from \cite{Volkovacos}.
As can be seen from Figure \ref{depth_angle}, there is no evident increase of 
the deflection of the data points from
the best fit predictions ($R_c=0$) with the increase of depth
at large $\cos \theta$ as it should be if the significant prompt
muon flux is present. The deepest depth bin is the exception. However,
due to small statistics, the data at very large depth do not affect much
the total value of $\chi^2$.

If the formula from \cite{Volkovasp} is used for the muon
spectrum at sea level instead of eq. (\ref{Gaisser spectrum}), the
best fit values of $\gamma$ will be decreased by 0.04-0.05 and will be
in agreement with the previously published values for single muons
\cite{LVD,LVDR}
analysed using the formula from \cite{Volkovasp}. This difference,
being comparable with our total error, is due to the factor which
is present in the formula from \cite{Volkovasp} and
takes into account the rise of hadron--nucleus cross-section 
at high energies. This factor appears in the
calculation \cite{Volkovasp} if
the rise of the total hadron--nucleus cross-section with energy is
due to the rise of the differential cross-section in the central
region, while the scaling is conserved in the fragmentation
region. This factor makes the muon energy
spectrum steeper and the difference in the power index
of muon spectrum is about 0.04-0.05. 

Similar analysis performed for single muons shows no evidence for
prompt muon flux, too. We found the same values of power index and
upper limit to the prompt muon flux, while the absolute intensity is 10\%
smaller.

\section { Discussion }

\indent From the analysis of the depth-angular and depth
distributions of muon intensities measured by LVD the following
estimates of the parameters of the muon spectrum at the sea level
have been obtained: $A=1.84 \pm 0.31$, $\gamma=2.77 \pm 0.02$
(68$\%$ C.L.), $R_c \leq 2 \cdot 10^{-3}$ (95$\%$ C.L.).
The errors include both statistical and systematic errors with the
systematic error dominating. The systematic error takes into
account the possible uncertainties in the depth and local density,
which have been estimated from the difference between the measured
and predicted intensities for all angular and depth bins. 
The uncertainties of rock thickness and local density both result in the
uncertainty of the column density and, hence, in the uncertainty of the muon
flux. The distribution of fractional differences between measured and
predicted intensities has been found to be close to gaussian with a standard
deviation of about 0.04. This value has been assumed as a systematic error
of muon intensity due to the column density uncertainty. This value is
equivalent to the column density error of about 1$\%$ at a depth of 3 km w.e.
It is obvious that the systematic error is more important at small depth where
the statistics is high and statistical error is negligibly small.
An additional systematic error due to the uncertainties of the
cross-sections of muon interactions used to simulate the muon
survival probabilities should be included. According to
the discussion in \cite{MUSIC} we estimate the total
uncertainty in $\gamma$ as 0.05 and in $A$ as 0.5.
The uncertainty in the cross-sections, however,
does not affect the upper limit to $R_c$.
To check this we have fitted LVD data with the intensities calculated with
muon bremsstrahlung cross-section from \cite{BBb} and obtained the 
following results:
$A=1.86 \pm 0.32$, $\gamma=2.78 \pm 0.02$
(68$\%$ C.L.), $R_c \leq 2 \cdot 10^{-3}$ (95$\%$ C.L.).
The muon bremsstrahlung 
cross-section from \cite{BBb} is a little smaller than that from 
\cite{KKP}. This makes
the muon 'depth-intensity' curve (with fixed $A$, $\gamma$ and $R_c$) 
flatter. This is
compensated in the data analysis by the increase of $\gamma$. But the shape 
of the calculated angular distribution of muon intensities at any fixed depth, 
used to extract the value of $R_c$, is not changed and, hence, the limit on 
ratio of prompt muon flux to that of pions remains unchanged. However, 
the absolute value of prompt muon flux (or its limit) varies with the 
muon cross-sections used since the flux depends also on normalization 
constant, $A$, and power index, $\gamma$ (see eq. (\ref{Gaisser spectrum})).

The value of $\gamma$ obtained with
LVD data is in reasonable agreement with the results of many other
surface and underground experiments (see, for example,
\cite{MSU,DEIS,MUTRON,ASD,MIPhI,NUSEX,MACROc}). However, the results
obtained in the experiments which used the 
indirect method
of the measurement of the muon spectrum, in particular,
the measurement of the depth--intensity curve, are strongly
affected by the muon interaction cross-sections and the
algorithm applied to calculate the muon intensities. We have
used the most accurate cross-sections, known at present, and
the algorithm which allows us to calculate the muon intensities
with an accuracy of $1\%$ for a given set of muon interaction
cross-sections and for homogeneous medium.
The algorithm can influence strongly the calculated muon
intensities and, then, the final results (for a discussion 
see, for example \cite{MUSIC}). Thus, the observed
agreement (or disagreement) in the value of $\gamma$
does not mean the agreement (or disagreement) in the data
themselves.

The conservative upper limit to the fraction of prompt
muons, obtained with the LVD data ($R_c<2 \cdot 10^{-3}$),
even in the simple assumption that the power index of the prompt
muon spectrum, $\gamma_c$, is equal to that of primaries,
$\gamma$, rules out many models of the prompt muon
production, which predict a fraction of prompt muons more than
$2 \cdot 10^{-3}$. To make this conclusion more reliable we have
carried out the analysis of the depth -- angular distribution of
muon intensity using the prompt muon spectra predicted by
different models (without a constant term $R_c$). We conclude
that the LVD data contradict the predictions of model 1
 \cite{Elbert}, model II \cite{Inazawa} and model A 
\cite{Zas}. The predictions of the model 3 \cite{Elbert}, model I
\cite{Inazawa}, models B, C \cite{Zas}, recombination
quark-parton model (RQPM) \cite{Bugaev} and model by
\cite{Volkovac} are comparable with the LVD upper limit, and
these models
cannot be ruled out. At the same time the LVD result favours
the models of charm production based on QGSM (see, for example,
\cite{Bugaev}) and the dual parton model \cite{Forti},
which predict low prompt muon flux.

The upper limit (95$\%$ C.L.) obtained with the LVD data
is lower than the value of $R_c$ found in the MSU experiment
($R_c=(2.6 \pm 0.8) \cdot 10^{-3}$ at $E_{\mu 0}=5$ TeV
\cite{MSU}). The LVD upper limit does not contradict
the values of prompt muon flux, obtained in
Baksan \cite{Baksanc} and KGF \cite{KGF}
underground experiments. Our result agrees with
that of NUSEX \cite{NUSEXc}
which did not reveal any deviation from the
angular distribution expected for conventional muons.

We point out that the LVD sensitivity to the prompt muon
flux is restricted mainly by the systematic uncertainties
connected with the uncertainties of the slant depth and local
density fluctuations
and the differences in the theoretical shape of the muon
underground intensities.

\section { Conclusions }

The analysis of the depth--angular
distribution of muon intensity measured by LVD
in the depth range 3000-10000 hg/cm$^2$ has been done. The
parameters of the muon energy spectrum at the sea level have
been obtained (see eq. (\ref{Gaisser spectrum})):
$A=1.8 \pm 0.5$, $\gamma=2.77 \pm 0.05$
and $R_c<2 \cdot 10^{-3}$ (95$\%$ C.L.). The errors include
both statistical and systematic uncertainties. The upper limit
to the fraction of prompt muons, $R_c$, favours the models
of charm production based on QGSM \cite{Bugaev} and the dual parton
model \cite{Forti}, and it rules out several models which predict
a high prompt muon flux.
Similar analysis performed for single muon events revealed the same values
of power index and upper limit to the fraction of prompt muons, while
the normalization constant is 10\% smaller.

\section {Acknowledgements}

We wish to thank the staff of the Gran Sasso Laboratory
for their aid and collaboration. This work is supported by the
Italian Institute for Nuclear Phy\-sics (INFN) and in part by the
Italian Ministry of University and Scientific-Technological
Research (MURST), the Russian Ministry of Science and Technologies,
the Russian Foundation of Basic Research
(grant 96-02-19007), the US Department of Energy, the US National
Science Foundation, the State of Texas under its TATRP program,
and Brown University.

\pagebreak

\pagebreak

\pagebreak

\end{document}